# Design and Implementation of A Network Security Management System

Zhiyong Shan[1], Bin Liao[2]

[1] School of Information, Renmin University of China, zyshan2003@hotmail.com

[2] School of Electric and Electronic Engineering, North China Electric Power University

**Abstract**: In recent years, the emerged network worms and attacks have distributive characteristic, which can spread globally in a very short time. Security management crossing network to co-defense network-wide attacks and improve efficiency of security administration is urgently needed. This paper proposes a hierarchical distributed network security management system (HD-NSMS), which can centrally manage security across networks. First describes the system in macrostructure and microstructure; then discusses three key problems when building HD-NSMS: device model, alert mechanism and emergency response mechanism; at last, describes the implementation of HD-NSMS. The paper is valuable for implementing NSMS in that it derives from a practical network security management system (NSMS).

## 1 Introduction

The earliest research on NSMS can be pursued back to the Europe project SAMSON [1] in 1992, which performs the security management by integrating interfaces of CMIP and SNMP. Scholars in the world discussed widely on NSMS in each subsequent year. In 1997, F. Stamatelopoulos etc. implemented an integrated network security management system on UNIX machine by adopting SNMP protocol and using architecture of Manager/Agent [2]. In 1998, Philip C. Hyland etc. proposed a three-stage theory for network security management and built a security management framework named CSSA [3]. In 1999, Soon Choul Kim etc. built a NSMS based on architecture of Client/Server [4], which collects security information from each computer, analysis them and displays results to users. Between 2000 and 2002, K. Boudaoud etc. discussed NSMS [5] from perspective of multi-agent, and built a prototype. But its main goal is to promote the ability of intrusion detection. In 2005, J. Dawkins etc. built a prototype of NSMS [6], which identifies, tracks and analyzes security events based on security information gotten from other detecting and scanning tools, and displays results to users in graph style.

Above researches focus on single network security management, most of them didn't involve cross network security management. However, requirement of security management crossing network is becoming more and more urgent lately. Because of that most of worms emerged in recent years have distributive characteristic, which can spread to worldwide in a very short time, and maybe derive from different networks and districts, these require extensive coordination in the procedure of tracing and defense. Cross network security management can meet the requirement well for that it is effective in defending network-wide attacks and managing security by the means of integrated management and cross-network cooperation when handling events, policies and vulnerabilities.

In the period of 2004~2006, we implemented a HD-NSMS that can execute cross network management, and mainly included five main functions: status monitor, event management, policy management, vulnerability management and security situation evaluation. On basis of the practical research and development, section 2 of the paper describes the system in macrostructure and microstructure; section 3 discusses three key problems when building HD-NSMS: device model, alert mechanism and emergency response mechanism; section 4 describes the implementation of HD-NSMS.

## 2 Architecture

HD-NSMS is a complicated system, whose architecture has to be described from both macro level and micro level.

### 2.1 Macro Architecture

HD-NSMS aims at protecting network of large-scale organization, which has three characteristics in general: hierarchical, distributed and dendriform, it is shown in Figure 1. Correspondingly, HD-NSMS should have consistent macro architecture with organizational network structure in macro level. Therefore, HD-NSMS is composed of a number of nodes. Each node is named as "security management node"(SMN), which is responsible for centralized security management in local network and its descendent networks. Corresponding to organization's network, all of the SMNs construct a tree like structure.

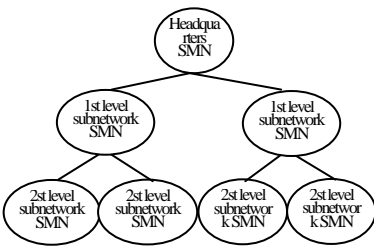

**Fig.1.** Network of large-scale organization

As there are notable differences between managing device and SMN, SMN can be divided into two classes: Intranet SMN (ISMN) and Extranet SMN (ESMN). ISMN manage device, and ESMN manage SMN. Hence, two kinds of macro architecture can be built: Macro Architecture combining ISMN and ESMN (see Figure 2), Macro Architecture separating ISMN and ESMN (see Figure 3). Device in this paper means security related equipments, hosts or software. Devices can be divided into three classes: security device, network device and important hosts.

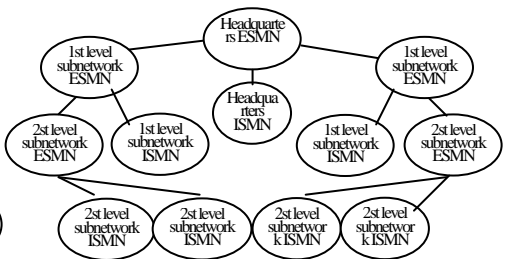

**Fig.2.** Macro Architecture combining ISMN and ESMN

**Fig.3.** Macro Architecture separating ISMN and ESMN

Under the macro architecture separating ISMN and ESMN, ISMN is responsible for the tasks related to device, such as monitoring running status of device, receiving and handling security events from device, acquiring security policy from device and enforcing policy to device. ESMN can fulfill security management in a larger network under the support of lower ISMN and ESMN, such as analyzing security events, computing security situation, and distributing security policies and patches etc. Although this architecture divides works of ISMN and ESMN clearly, the cost of developing and maintaining such kind of management system is high. Besides that, the whole system will be complicated for that it comprises two types of SMN.

Under the merged architecture, there is only one type of SMN, which manages both device and SMN. As this architecture regards SMN as a special device, the structure of the HD-NSMS looks simple, and the cost of development and maintenance is low. Therefore, we choose the architecture combining ISMN and ESMN for our HD-NSMS.

| User Interface |
| --- |
| Function Logic |
| Device Logic |
| Common Communication |
| Device Interface |

**Fig.4.** Five Layers of Micro Architecture

### 2.2 Micro Architecture

Diversity is an important characteristic of NSMS, which includes platform diversity, device type diversity, manufacturer diversity and information structure diversity. How to merge all of these together and at the same time guarantee high performance, reliability, security and scalability of the system is not easy to resolve. The best way to resolve such problem is adopting method of layered abstraction. So, we abstract micro architecture of HD-NSMS into five-layer depicted in Figure 4.

The ground layer is DIL (Device Interface Layer), which acquires and configures directly various data or files on device, masks differences and details of device operations and data derived from different manufacturers, and provides unified device operations and data interfaces for upper layer.

The second layer is CCL (Common Communication Layer), which is responsible for communication between agents and SMN, between user console and SMN, and between SMN and SMN. This kind of communication includes communication in local network or across wide area network. Besides that, encryption, certification, speed control, reliability, priority control and time synchronization all are involved in. Providing unified communication interface to SMN, agent and user console is also concerned.

The third layer is DLL (Device Logic Layer), which is responsible for maintaining device topology in SMN, masking some details such as device type or location etc, and providing more abstract device operations for upper layer.

The fourth layer is FLL (Function Logic Layer), which helps to implement SMN functions of device status monitoring, events management, policy management, vulnerability management and security situation evaluation.

The top layer is UIL (User Interface Level), which runs on user console. It interacts with users directly in graph style, and sends user commands and data to FLL.

Table 1 describes the abstraction of each layer.

**Table 1.** Five-layer structure

| Layer | Function | Main operations for upper layer | Data for upper layer |
|---|---|---|---|
| UIL | Interacts with users | Graph Interface | Graph Interface |
| FLL | Implements device status monitoring, events management, policy management and security situation evaluation | Send device status; send device event; query event; read/write device security policy; dispatch/receive patch; scan vulnerability; evaluate security situation; handle security event. | Command parameters |
| DLL | Maintains device topology | Build, assemble, disassemble, expand, and shrink device topology; read, write, search, add, delete, and modify device. | Device No. |
| CCL | Communicates between agent and SMN, user console and SMN, SMN and SMN | Send data; Receive data | Communication data（decided by communication protocol） |
| DIL | Manages device directly | Read device status; read device log, receive device event; read/write device security policy; write device patch. | Device running status, device events or log, device security policy, device patches |

Building NSMS based on the layered-abstraction idea has merits of that it can merge all devices of different types or from different manufactures together conveniently. It is also convenient to add new functions into NSMS. Moreover, the architecture is clear and east to maintain.

The micro-architecture is built as Figure 5, which is divided into four independent parts: Security Management Node (SMN), Device Agent (DA), User Console (UC) and Common Communication Server (CCS). These four parts correspond to above five-layer structure.

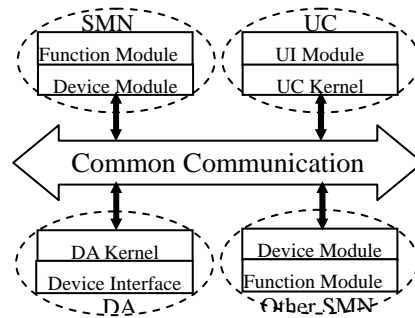

**Fig.5.** Micro Architecture of HD-NSMS

DA is composed of Device Interface module and DA Kernel module. Both of them belong to DIL. Device Interface module communicates directly with device through several kinds of protocols, which includes SNMP, CMIP, syslog, manufacturer specific protocols, customed protocols or directly accessing. By these protocols, the module can complete operations of getting or setting device status, getting or setting device policy, etc. DA Kernel receives and explains commands from SMN, and performs required operations on device through device interface module. At the same time, it receives alerts and status information from device, transforms them into uniform formats and sends them to SMN through CCS.

Common Communication is composed of communication library and CCS, and belongs to CCL. Communication library is attached to SMN, DA and UC, implements communication protocol stack, multi-threads parallelization and multi-level cache etc. CCS is an independently running program, mainly responsible for encryption/decryption, cross-network speed control, asynchronous storage and routing of communication.

SMN is the kernel of HD-NSMS and is composed of Device Module and Function Module. Device Module belongs to DLL, it maintains devices' logical topology, attaches device attributes for information sending out, explains device attributes for information receiving in. Function module belongs to FLL, performs functions of SMN, and manages parallel asynchronous information. Except of managing devices in local network, SMN also communicates with other SMNs, regards descendent SMN as its own devices, and regards itself as superior SMN's device.

UC Kernel and UI Module belong to UIL. They are responsible for user interaction. UI Module is a graphical interface. UC Kernel creates and sends user commands, receives and explains alert and status information, and maintains device topology in user view.

## 3 Key Issues

### 3.1 Device Model

Device is the key object of HD-NSMS, and the five functions of HD-NSMS all are relative to device, so building reasonable device model is very important. The process of building device model is divided into three steps of addressing, abstraction and organization.

### 3.1.1 Device Addressing

HD-NSMS is a typical large-scale distributed communication system. Network communication is required between all device agents and SMN. Therefore, uniform addressing is necessary for identifying each communication node uniquely in whole HD-NSMS.

The addressing method of HD-NSMS should satisfy two requirements: uniqueness and locatability. Uniqueness means identifying a node uniquely in the macro tree structure of HD-NSMS, even if two nodes are located in the same LAN or on the same machine; Locatability means the logical location of a node in the macro structure tree of HD-NSMS can be ascertained. In other words, the level, sub-tree or number of any node can be ascertained.

For identifying a communication node, IP address is a typical example, which can identify a machine on the Internet uniquely. But, if HD-NSMS uses IP for identification, there will be some problems difficult to resolve: How to ascertain the logical position of a node in HD-NSMS' macro structure tree by using IP address? How to differentiate each node while several communication nodes share one IP address? Hence, IP address isn't appropriate for HD-NSMS addressing.

We designed a new addressing method. It assumes $n$ and $d$ as depth and degree of HD-NSMS'

macro architecture tree, represents address of a node as following string:

$$A1.A2.A3....An \qquad 0 \leq A_i \leq d \quad 1 \leq i \leq n$$

The node address string is composed of several segments separated by dots. The amount of segments is equal to the depth or total levels of macro structure tree. Each segment describes number of the node under its parent node. For example, there is a node that has address as 1.2.4.0.0.0.0.0, which means the HD-NSMS has eight levels in total and this node is a 3rd level node. The number of this node in level 1 is 1, level 2 is 2, and level 3 is 4. The numbers below level 3 all are zero, which indicates level of the node is 3. This addressing method meets above two requirements well for that an address can represent a HD-NSMS node's logical position and identify it uniquely.

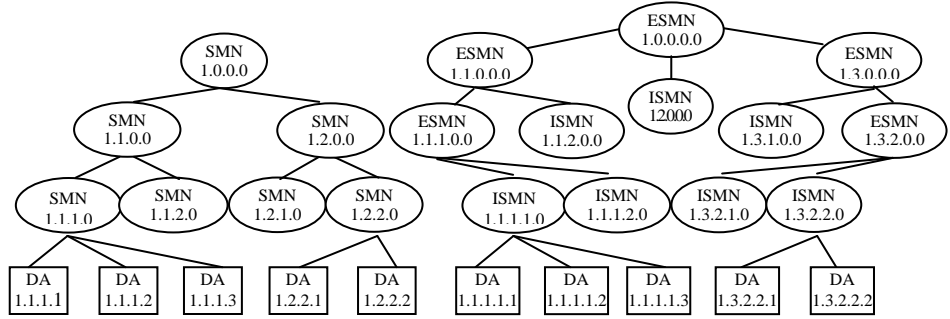

**Fig.6.** Addressing examples for combined architecture and separated architecture

We illustrate HD-NSMS addressing method by using the network structure in Figure 1. Left part of figure 6 for merged macro architecture, right part for separated macro architecture.

Obviously, all nodes in HD-NSMS construct a tree. Each node in the tree represents a device and has an address. We name such tree as **Addressed Device Tree**. In the Addressed Device Tree, there are only two kinds of nodes: SMN node and Device node. Only SMN node can be a non-leave node.

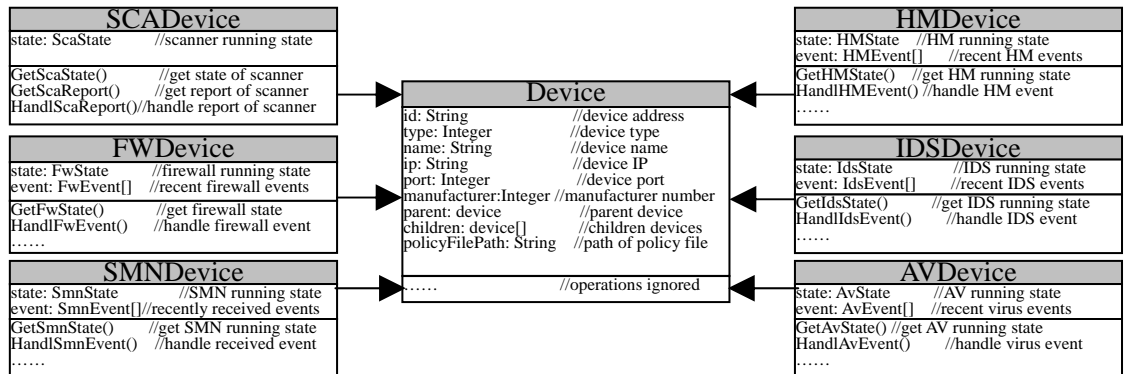

**Fig.7.** Static abstraction of devices

For any node in Addressed Device Tree, the algorithm complexity of searching a node from the root is O (n), which n is depth of the node in the tree. In the case of numberless tree, the algorithm complexity should be O (n*d), which d is degree of the node. So Addressed Device Tree is efficient on performing operations of search, add and delete.

### 3.1.2 Device Abstraction

We abstract device both statically and dynamically. Static abstraction regards security device, network device, host machine and SMN as a uniform device object which has attributes of device type, operations and properties. Device object is the kernel data structure of the whole HD-NSMS, we describes it by UML in Figure 7. Firstly, we define a super class named Device, which includes basic information about device and related operations; then, define each specified device object which includes attributes of device' running status, latest events list and corresponding handling functions. Specific device object involves FWDevice, IDSDevice, AVDevice, SCADevice, HMDevice and SMNDevice.

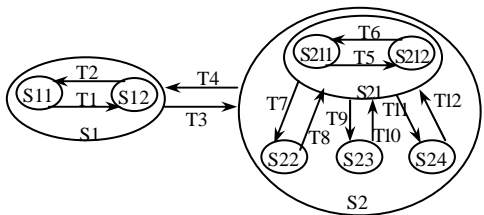

**Fig.8.** Dynamic abstraction of devices

In addition, for abstracting device dynamically, we describe a state transition diagram that contains ten statuses and twelve transfer conditions as shown in Figure 8. Ten status that pictured as nested circles include offline (S1); network disconnected (S11); device can't arrive (S12); online (S2); waiting (S21); running normally (S211); running abnormally (S212); handling alert (S22); handling policy (S23); handling vulnerability (S24). Twelve transfer conditions include receiving network test package (T1); overtime when waiting network test package (T2); receiving device state package (T3); overtime when waiting device state package (T4); device abnormity of CPU, memory, disk, network flow or key process (T5); device normality of CPU, memory, disk, network flow or key process (T6); receiving device alert (T7); finishing handling alert (T8); receiving policy handling command (T9); finishing handling policy command (T10); receiving vulnerability handling command (T11); finishing handling vulnerability command (T12).

### 3.1.3 Device Organization

Device organization should reflect topology of physical devices in reality, and convenient for perform any computation. Besides, device organization should be displayed in visualized way for users to understand and operate. Therefore, we propose a three mapping model for device organization: physical view, virtual view and user view.

➢ Physical View: a device topology that describes how the physical devices are connected in reality.

➢ Virtual View: a mapping of the physical view in memory. It also can be regarded as a topology of abstracted device objects. A device in virtual view is a logical abstraction of the corresponding physical device. Virtual view is the kernel of three layers and computational foundation for five functions of HD-NSMS.

➢ User View: a mapping of virtual view on user interface. It also shows the result of user operation. Devices in user view are different from those in virtual view. User View describes position, size, icon, alert and sounds of device objects in user interface. It not only can map virtual view, but also can modify virtual view in return.

Essentially, virtual view and user view are Addressed Device Tree composed of abstract devices. Virtual view and user view have to change dynamically for mapping updates of physical device or making user operation conveniently. Possible changes in virtual view include: building, assembling and disassembling tree, adding and deleting device, modifying and switching status. Possible changes in user view include: building, destroying, assembling, disassembling, expanding and shrinking tree; adding and deleting device; modifying and switching status. Before introducing the algorithm of dynamical changing, two concepts need to be defined:

**Device Tree's Embedding Structure**: a data structure for encapsulating Addressed Device Tree, used in transmitting the Addressed Device Tree between virtual view and user view, and between upper SMN and lower SMN. Device tree's embedding structure and Addressed Device Tree can convert reciprocally. It describes each device in following format:

[DeviceId:DeviceState:Child1:Child2:Child3……]

Then, the device tree in left of Figure 6 can be represented as:

[1.0.0.0:state:[1.1.0.0:state:[1.1.1.0:state:[1.1.1.1:state]:[1.1.1.2:state]:[ 1.1.1.3:state]]:[ 1.1.2.0:state]]:[ 1.2.0.0:state:[ 1.2.1.0:state]:[ 1.2.2.0:state:[ 1.2.2.1:state]:[ 1.2.2.2:state]]]]

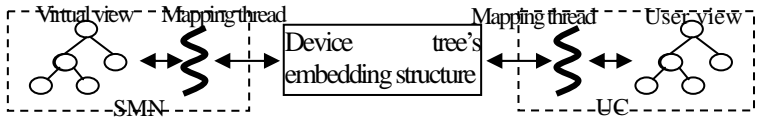

**Fig.9.** View mapping mechanism

**View Mapping Thread**: a daemon thread running on SMN and UC. Under the help of Device Tree's Embedding Structure, it may implement mapping between virtual view and user view, which is illustrated in Figure 9.

For an administrator of SMN, he/she only knows number, status and other information of direct subordinate devices. It is difficult for him/her to fill the information of indirect subordinate devices, because indirect subordinate devices maybe locate in remote place geographically. Therefore, we introduce assembling and disassembling algorithms to update automatically of subordinate device tree.

Assembling algorithm can assemble the device tree's embedding structure under the main tree and update the mapping onto user view. The assembling algorithm is given out as Figure 10. If subordinate SMN doesn't send information for a long time, it is considered offline or network failure. At this time, its subordinate sub-tree should be disassembled. The disassembling algorithm is given out as Figure 11.

Assembling algorithm:
Parameter: device tree's embedding structure
1) call algorithm of building tree to transform device tree's embedding structure to a subtree preparing for assembling;
2) search in main tree the assembling node that is the same as root node of the subtree;
3) add root node of the subtree into children device group of parent node of the assembling node;
4) set parent of the assembling node as parent of root node of the subtree;
5) delete the assembling node from its parent's children device group;
6) free the assembling node in virtual view;
7) send new device tree to user view by mapping thread;
8) find assembling node in main tree of the user view;
9) update state of the assembling node according to root node of the subtree;
10) by using an search algorithm of depth precedence, add every node of subtree circularly under the assembling node in user view;
11) update user view;

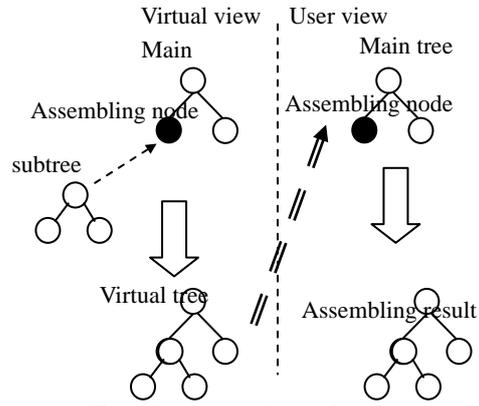

**Fig.10.** Assembling algorithm

Disassembling algorithm:
Parameter: disassembling node
1) find the disassembling node in main tree of virtual view;
2) modify state of the disassembling node;
3) by using an search algorithm of depth precedence, delete every descendent nodes of the disassembling node;
4) send new device tree to user view by mapping thread;
5) find the disassembling node in main tree of user view;
6) update state of the disassembling node in user view;
7) by using an search algorithm of depth precedence, delete circularly every descendent node of the disassembling node;
8) update user view;

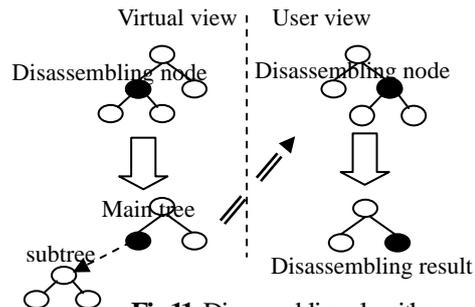

**Fig.11.** Disassembling algorithm

### 3.2 Alert Mechanism

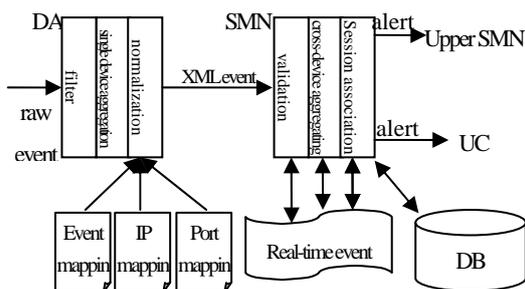

**Fig.12.** Alert mechanism

Objective of alert mechanism is integrating events and disclosing relations among events. In recent years, a lot of research has been done in this field [9~13], but most of them keep eyes on intrusion detection and doesn't involve relation analysis across different type of devices. Generally, they don't treat differently on events from different type of devices and seldom associate the events based on inner relationship of events from different type of devices. Cross-device event-association is an inevitable problem that has to be solved for building HD-NSMS. Our alert mechanism is described in Figure 12.

DA accomplishes single-device event handling. Filter module of DA removes useless events coming from Firewall, IDS and Host Monitor etc. Even in the case of no attacks happen useless events may be generated largely, paper [15] proves this, so filtering is necessary. Single device aggregation module merges repeated events, ports scanning events, host monitoring events and DOS events. The normalization module converts each device event into a uniform XML format satisfying the requirements of IDMEF，and prepares to cross-device aggregation.

SMN accomplishes cross-device event handling. Validation module of SMN rates normalized events that received from DA and filters the lowest events, rating event lies on value loss computed from value and vulnerabilities of information asset who is attack target of the event. Based on the event normalization, cross-device aggregating module aggregates events from multiple types of devices by uses a similarity function that has the form as in paper [16]. Session association module associates events belonged to the same network session into a session alert, and send the alert to UC.

Session alert means aggregation of all related events during one network connection. These events surely belong to one attack and have strong relation between each other. Session alert is composed of session information and event queue. Session information mainly describes the source and destination address, start and stop time of the session. Events queue means events sorted by time and is useful for administrator to analyze the attacker's intention.

The key to build session alert is finding the events of connection and disconnection. The events can be sort out from firewall logs as the reason of that firewall monitors all kinds of connections, which surely contains connections of attack. For connection-oriented protocols, we can easily get connection and disconnection events; for non-connection-oriented protocols, we also can get the events in that most firewalls support stateful-inspection technology who can treat non-connection-oriented protocol as having connection [14].

Each session alert has a session alert object in SMN memory, the object has an internal event queue, and the session alert is generated from the first non-connection event and destroyed when the connection is closed. At the same time, there is a global event queue for recording beginning time of each connection. Left of following describes session association algorithm, and right of following is an algorithm that deciding whether an event belongs to a session alert.

```
Input: event
Output: void
If(received event of building connection)
    save the event into the global event queue of building connection;
If(received event of non-connection){
    search all of session alert objects;
    If(corresponding session alert don't exist){
        search in the global event queue of building connection;
        If(corresponding event of building connection exist){
            build a new session alert object;
            initiate the alert object according to the event;
            build a new event queue;
            destroy the event of building connection;
        }else{
            regards the event as an independent session alert;
            send alert ending information to UC, database or upper SMN;
            clean the session alert and its event queue;
            return；
        }
    }
    append the event to event queue of the corresponding session alert;
    send the session alert to UC or database;
}
If(received event of disconnection){
    If(corresponding session alert exist){
        write time of connection ending into session alert;
        send alert ending information to UC, database or upper SMN;
        set a destroyng flag on session alert for destroying the alert after a while;
    }else{
        search in the global event queue of building connection;
        If(corresponding event of building connection exist)
            destroy the event of building connection;
    }
}
Return；
```

```
Input: event and session alert
Output:the event whether belongs to the session alert
If((source and target address of the event both is the same as the session alert)or(source address of the event is the same as target address of the session alert))
    If( event time is after connection building event of the session alert)
        If(event time is before disconnection event of the session alert)
            Return(the event belongs to the alert)
search in event queue of the session alert
if(source address of the event is the same as target address of certain event of the session alert)
    If(event time is after building connection event of the session alert)
        If(event time is before disconnection event of the session alert)
            Return(the event belong to the alert);
Return(the event doesn't belong to the alert);
```

Session alert supported event association mechanism has several advantages. First, it can complete association in real-time as a result of that association operation is done in memory and searching queue is short; Second, it uses network session as unit to cluster events instead of usual time window, because of that the events belonged to the same session have more tight relationships with each other than the events belonged to the same time window; Third, it is easy to find out in-progress attack for that the session alert not ended indicates a in-progress attack; Fourth, it is easy to linkage with firewall by session information.

**3.3 Emergency Response Mechanism**

At present, handling security emergency mostly depends on manual operation, so it imminently needs an automatic or semi-automatic mechanism. We design a semi-automatic emergency handling mechanism for HD-NSMS, and the mechanism synthesizes three factors of flow, counterplan and cooperation.

Flow means that the whole emergency response progress should comply with some normative flow. We design the emergency response flow in accordance with FCC drew Computer Security Incident Response Guide [8], which divides emergency response flow into six steps: Preparation, Identification, Containment, Eradication, Recovery and Follow-up. The steps of Preparation and Follow-up are not included in HD-NSMS by reason that these two steps are minor in handling emergency.

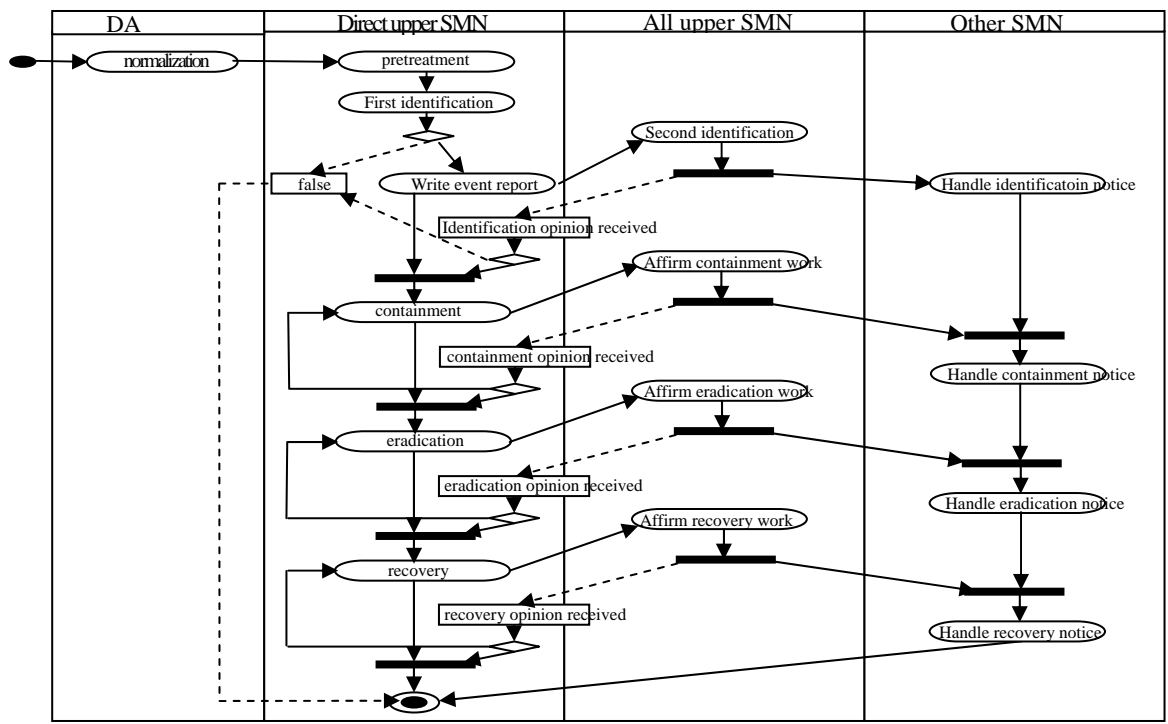

**Fig.13.** Cooperative emergency response in network-wide

Counterplan is a document that is written beforehand and saved in database, and it is used to guide event handling. Every event is associated with a counterplan. Counterplan consists of four sub-counterplans of identification, containment, eradication and recovery, which in accordance with the main four steps of emergency response flow. When administrator launches emergency response flow, HD-NSMS automatically sorts out corresponding counterplan and displays it to user interface.

Cooperation means that HD-NSMS provides a mechanism for entire network's SMN nodes to treat together important security event. For the local network that security incident occurred in, its SMN is the main participant for event handling. As the higher-level network administrator usually has better experience in security, he/she is regarded as the coordinator in whole handling progress. For wide range attacks or worms, HD-NSMS adopted other SMN into handling progress to implement cooperative defense. We use UML to describe the handling progress in Figure 13.

## 4 Implementation

From 2004 to 2006, based on the previously described system structure, device model, alert mechanism and emergency response mechanism, we achieved an HD-NSMS system. Figure 14 depicts the hierarchical subsystems dependent relationship by method of UML. It ignores tool classes, supporting function classes and thread classes. Its layers basically correspond to the modules in Figure 5 that depicts microstructure of HD-NSMS.

DAKernel Layer and DevInt Layer correspond to the "DA Kernel" and "Device Interface" modules in Figure 5. DAEvent module in DAKernel Layer is responsible for filtering, single-device aggregating and normalizing event.

CommCom Layer corresponds to the "Common Communication" module in Figure 5, and performs communications between SMNs and DAs.

DBInt Layer provides unified DB operations for other modules.

DevTopo Layer corresponds to the "Device Modules" in Figure 5. Five subsystems of IDSDevice, FWDevice, AVDevice, ScannerDevice and HMDevice manage respectively five types device of IDS, firewall, anti-virus, scanner and host-monitoring system. SMN is abstracted as device object and managed by SMNDevice module. DeviceManager is the external interface of DevTopo Layer, and implements maintenance operations on logic topology of devices.

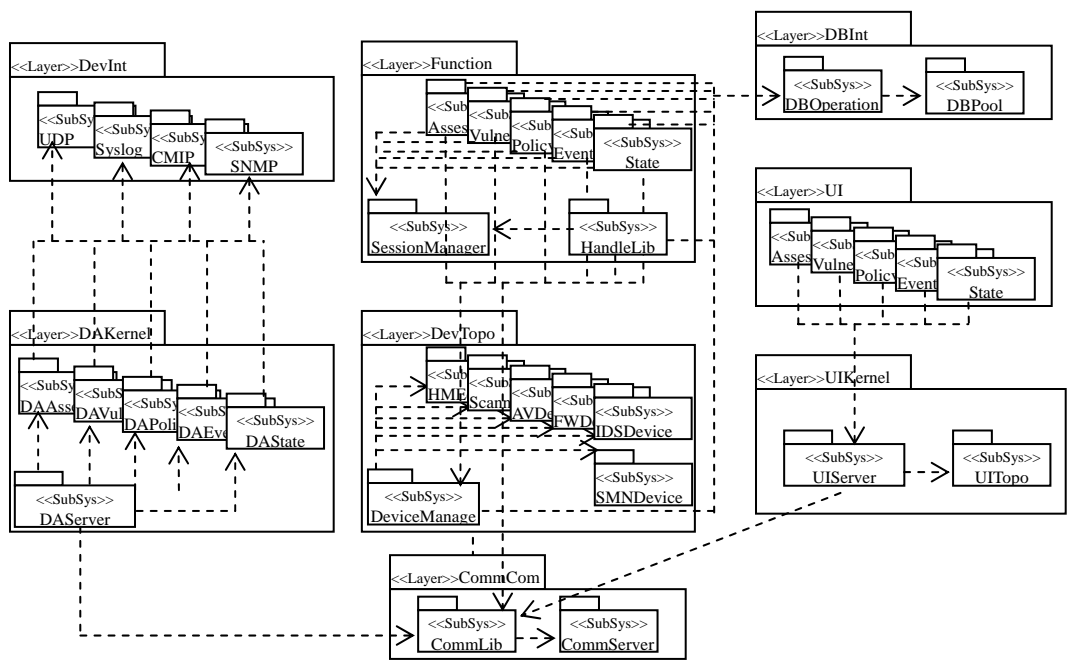

**Fig.14.** HD-NSMS implementation depicted by UML

Function layer corresponds to the "Function Modules" in Figure 5. Five subsystems of State, Event, Policy, Vulnerability and Assessment archive respectively five functions of HD-NSMS. HandleLib module handles asynchronously the input packages. SessionManager module manages session queue and runs session monitoring thread. Event module is responsible for verifying, cross-device aggregating, associating and handling events, and implements the emergency response mechanism.

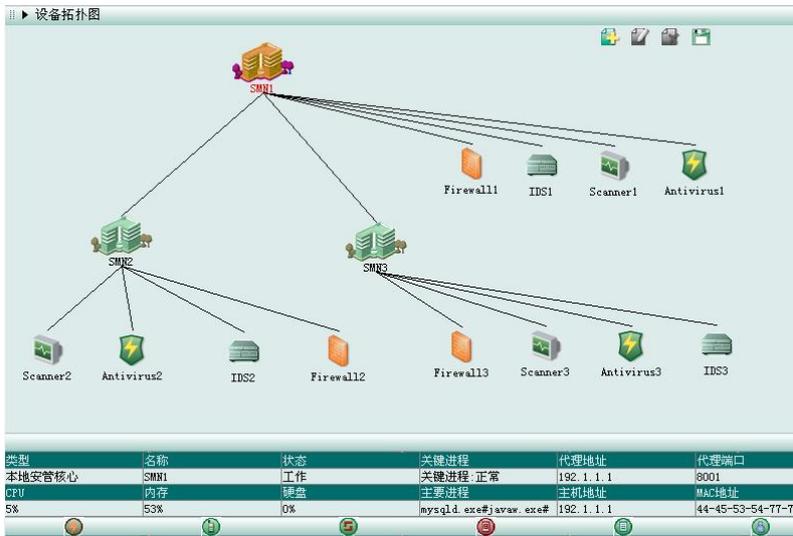

Fig.15. Devices topology in User Console

UI Layer and UIKernel Layer correspond to the "UI Modules" and "UC Kernel" module in Figure 5. Figure 15 shows a device topology in User Console of HD-NSMS.

## 5 Related Work

Compare to the related works [1~6], this paper innovatively builds a hierarchical distributed NSMS that can manage security in a Hierarchical distributed network. Though, there are several kinds of NSMS architectures, such as Manager/Agent [2], Client/Server [4] and Multi-Agent [5], they mainly focus on single-network security management, they can't manage policy, event or vulnerability in a cross-network environment. If having interest, please also read our other papers [32]-[71].

## 6 Conclusion

We promote architecture of HD-NSMS, and discuss three key problems in building HD-NSMS, which are device model, alert mechanism and emergency response. We describe the architecture from macrostructure and microstructure. Macrostructure has two kinds which both have advantages and disadvantages, one is ISMN and ESMN combined architecture; another is ISMN and ESMN separated architecture. Microstructure is divided into five layers: user interface, function logic, device logic, common communication and device interface. This layered architecture can screen differences among device types, manufacturers, or data structures, and provide a unified security management interface to administrator. Using this five layers architecture, we can build a HD-NSMS that has good performance, dependability and extensibility.

As device is kernel object of a NSMS, we build a device model from respects of identifying, abstracting and organizing. For identifying device, we design a device addressing method that codes device address in a string, which composed of several numbers and separated by dot. For abstracting device, we use UML to describe abstracted device objects, and depict state transition diagram of abstractive device object. For organizing device, we promote a three level mapping model that contains physical view, virtual view and user view, physical view is a topology graph of physical devices;

virtual view is mapped from physical view and exists in main memory; user view is mapped from virtual view and exists in GUI. In order to build the virtual view and user view of whole network in real time, we depict assembling algorithm that add devices of lower level network into the views, and depict disassembling algorithm that clean devices of lower level network from the views.

For alert mechanism, we promote a new event-association method based on session alert. We cluster all alerts that related to the same network connection into a session alert, because these events surely belong to the same attack, and can help administrator to analyze attack intent. The beginning and ending time of the network connection can be identified by the firewall event of building and destroying connection. Compare to paper [9] and [11], event-association method of this paper considers internal relations among events that generated from different types of devices, i.e. relation among firewall event and other device's event. Compare to paper [10], the associating method of this paper can do associating work in real time.

As currently people often do cross-network emergency response manually, we build a semi-automatic mechanism for cross-network emergency response. The mechanism considers three factors of emergency handing flow, counterplan and cooperative operation, can be used by administrators on different locations to cooperatively handle emergency.

## References

[1] S. Lechner, "SAMSON: Management of Security in Open Systems", Computer Communications, Sep 1994.
[2] F. Stamatelopoulos, G. Koutepas, B. Maglaris., "System Security Management via SNMP", Proceedings of the 4th HPOVUA Workshop on Network and Systems Management, Madrid, Spain, April 1997.
[3] Hyland P C. Concentric Supervision of Security Applications: A New Security Management Paradigm. In: Annual Computer Security Applications Conf. 1998
[4] Soon Choul Kim, Young Su Choi, Jin Wook Chung, "Study of security management system based on client/server model", ICC 1999 - IEEE International Conference on Communications, no. 1, June 1999 pp. 1403-1408.
[5] K. Boudaoud and C. McCathieNevile. An intelligent agent-based model for security management. Proc. of the Seventh IEEE Symposium on Computers and Communications, July 1-4 2002.
[6] J. Dawkins, K. Clark, G. Manes, and M. Papa. A Framework for Unified Network Security Management: Identifying and Tracking Security Threats on Converged Networks. Journal of Network and Systems Management, Vol. 13, No. 3, September 2005.
[7] Kienzle DM, Elder MC. Recent worms: A survey and trends. In: Staniford S, ed. Proc. of the ACM CCS Workshop on Rapid Malcode (WORM 2003). Washington, 2003.
[8] Federal Communications Commision, Computer Security Incident Response Guide, 2001.12. www.fcc.gov
[9] H. Debar and A. Wespi, "Aggregation and Alert Correlation of Intrusion Detection Alerts", Conference on Recent Advances in Intrusion Detection (RAID 2001), pp.85-103, Oct., 2001.
[10] P. Ning, Y. Cui, and D. R. f and D. Xu. Techniques and tools for analyzing intrusion alerts. ACM Transactions on Information and System Security (TISSEC), 7(2):274–318, May. 2004.
[11] Porras, P., Fong, M., and Valdes, A. 2002. A mission-impact-based approach to INFOSEC alarm correlation. In Proceedings of the 5th International Symposium on Recent Advances in Intrusion Detection (RAID 2002). 95–114.
[12] F. Cuppens. Managing Alerts in a Multi- Intrusion Detection Environment. Proceedings 17th Computer Security Applications Conference, New Orleans, LA, December2001.
[13] Cuppens and Miege 2002 CUPPENS, F. AND MIEGE. A. 2002. Alert correlation in a cooperative intrusion detection framework[A]. In : Proceedings of the 2002 IEEE Symposium on Security and Privacy[C], 2002.
[14] Check Point Software Technologies Ltd.http://www.checkpoint.com/products/downloads/Stateful_Inspection.pdf. 2005
[15] Tobias Chyssler, Simin Nadjm-Tehrani, Stefan Burschka, Kalle Burbeck: Alarm Reduction and Correlation in Defence of IP Networks. WETICE 2004: 229-234
[16] VALDES A, SKINNER K. Probabilistic alert correlation. In Proceedings of the 4th International Symposium on Recent Advances in Intrusion Detection(RAID 2001).
[17] Zhiyong Shan, Tanzirul Azim, Iulian Neamtiu. Finding Resume and Restart Errors in Android Applications. ACM Conference on Object-Oriented Programming, Systems, Languages & Applications (OOPSLA'16), November 2016. Accepted.
[18] Zhiyong Shan, I. Neamtiu, Z. Qian and D. Torrieri, "Proactive restart as cyber maneuver for Android," Military Communications Conference, MILCOM 2015 - 2015 IEEE, Tampa, FL, 2015, pp. 19-24.
[19] Jin, Xinxin, Soyeon Park, Tianwei Sheng, Rishan Chen, Zhiyong Shan, and Yuanyuan Zhou. "FTXen: Making hypervisor resilient to hardware faults on relaxed cores." In 2015 IEEE 21st International Symposium on High Performance Computer Architecture (HPCA'15), pp. 451-462. IEEE, 2015.
[20] Zhiyong Shan, Xin Wang, Tzi-cker Chiueh: Shuttle: Facilitating Inter-Application Interactions for OS-Level Virtualization. IEEE Trans. Computers 63(5): 1220-1233 (2014)
[21] Zhiyong Shan, Xin Wang, Tzi-cker Chiueh: Growing Grapes in Your Computer to Defend Against Malware. IEEE Transactions on Information Forensics and Security 9(2): 196-207 (2014)
[22] Zhiyong Shan, Xin Wang, Tzi-cker Chiueh: Malware Clearance for Secure Commitment of OS-Level Virtual Machines. IEEE Transactions on Dependable and Secure Computing. 10(2): 70-83 (2013)
[23] Zhiyong Shan, Xin Wang, Tzi-cker Chiueh: Enforcing Mandatory Access Control in Commodity OS to Disable Malware. IEEE Transactions on Dependable and Secure Computing 9(4): 541-555 (2012)
[24] Zhiyong Shan, Xin Wang, Tzi-cker Chiueh, Xiaofeng Meng: Facilitating inter-application interactions for OS-level virtualization. In Proceedings of the 8th ACM Annual International Conference on Virtual Execution Environments (VEE'12), 75-86
[25] Zhiyong Shan, Xin Wang, Tzi-cker Chiueh, and Xiaofeng Meng. "Safe side effects commitment for OS-level virtualization." In Proceedings of the 8th ACM international conference on Autonomic computing (ICAC'11), pp. 111-120. ACM, 2011.
[26] Zhiyong Shan, Xin Wang, and Tzi-cker Chiueh. 2011. Tracer: enforcing mandatory access control in commodity OS with the support of light-weight intrusion detection and tracing. In Proceedings of the 6th ACM Symposium on Information, Computer and Communications Security (ASIACCS '11). ACM, New York, NY, USA, 135-144. (full paper acceptance rate 16%)
[27] Shan, Zhiyong, Tzi-cker Chiueh, and Xin Wang. "Virtualizing system and ordinary services in Windows-based OS-level virtual machines." In Proceedings of the 2011 ACM Symposium on Applied Computing, pp. 579-583. ACM, 2011.
[28] Shan, Zhiyong, Yang Yu, and Tzi-cker Chiueh. "Confining windows inter-process communications for OS-level virtual machine." In Proceedings of the 1st EuroSys Workshop on Virtualization Technology for Dependable Systems, pp. 30-35. ACM, 2009.
[29] Shan, Zhiyong. "Compatible and Usable Mandatory Access Control for Good-enough OS Security." In Electronic Commerce and Security, 2009. ISECS'09. Second International Symposium on, vol. 1, pp. 246-250. IEEE, 2009.
[30] Xiao Li, Wenchang Shi, Zhaohui Liang, Bin Liang, Zhiyong Shan. Operating System Mechanisms for TPM-Based Lifetime Measurement of Process Integrity. Proceedings of the IEEE 6th International Conference on Mobile Adhoc and Sensor Systems (MASS 2009), Oct., 2009, Macau SAR, P.R.China, IEEE Computer Society. pp. 783–789.
[31] Xiao Li, Wenchang Shi, Zhaohui Liang, Bin Liang, Zhiyong Shan. Design of an Architecture for Process Runtime Integrity Measurement. Microelectronics & Computer, Vol.26, No.9, Sep 2009:183~186. (in Chinese)
[32] Zhiyong Shan, Wenchang Shi. "STBAC: A New Access Control Model for Operating System". Journal of Computer Research and Development, Vol.45, No.5, 2008: 758~764.(in Chinese)
[33] Liang Wang, Yuepeng Li, Zhiyong Shan, Xiaoping Yang. Dependency Graph based Intrusion Detection. National Computer Security Conference, 2008. (in Chinese)
[34] Zhiyong Shan, Wenchang Shi. "An Access Control Model for Enhancing Survivability". Computer Engineering and Applications, 2008.12. (in Chinese)
[35] Shi Wen Chang, Shan Zhi-Yong. "A Method for Studying Fine Grained Trust Chain on Operating System". Computer Science, Vol.35, No.9, 2008, 35(9):1-4. (in Chinese)
[36] Liang B, Liu H, Shi W, Shan Z. Automatic detection of integer sign vulnerabilities. In International Conference on Information and Automation, ICIA 2008. (pp. 1204-1209). IEEE.
[37] Zhiyong Shan, Qiuyue Wang, Xiaofeng Meng. "An OS Security Protection Model for Defeating Attacks from Network", the Third International Conference on Information Systems Security (ICISS 2007), 25-36.
[38] Zhiyong Shan, "A Security Administration Framework for Security OS Following CC", Computer Engineering, 2007.5, 33(09):151-163. (in Chinese)
[39] Shan Zhiyong, "Research on Framework for Multi-policy", Computer Engineering, 2007.5, 33(09):148-160. (in Chinese)
[40] Zhiyong Shan, Shi Wenchang, Liao Bin. "Research on the Hierarchical and Distributed Network Security Management System". Computer Engineering and Applications, 2007.3, 43(2):20-24. (in Chinese)
[41] Zhiyong Shan, "An Architecture for the Hierarchical and Distributed Network Security Management System", Computer Engineering and Designing, 2007.7, 28(14):3316-3320. (in Chinese)
[42] Shan Zhi Yong, Sun Yu Fang, "Study and Implementation of Double-Levels-Cache GFAC", Chinese Journal of Computers, Nov, 2004, 27(11):1576-1584. (in Chinese)
[43] Zhiyong Shan, Yufang Sun, "An Operating System Oriented RBAC Model and Its Implementation", Journal of Computer Research and Development, Feb, 2004, 41(2):287-298. (in Chinese)
[44] Zhiyong Shan, Yufang Sun, "A Study of Extending Generalized Framework for Access Control", Journal of Computer Research and Development, Feb, 2003, 40(2):228-234. (in Chinese)
[45] Shan Zhi Yong, Sun Yu Fang, "A Study of Generalized Environment-Adaptable Multi-Policies Supporting Framework", Journal of Computer Research and Development, Feb, 2003, 40(2):235-244. (in Chinese)
[46] Shan Zhiyong, Research on the Framework for Multi-Policies and Practice in Secure Operation System. Phd Thesis, Institute of Software, Chinese Academy of Science 2003. (in Chinese)
[47] Shan Zhi Yong, Sun Yu Fang, "A Study of Security Attributes Immediate Revocation in Secure OS", Journal of Computer Research and Development, Dec, 2002, 39(12):1681-1688. (in Chinese)
[48] Shi Wen Chang, Sun Yu Fang, Liang Hong Liang, Zhang Xiang Feng, Zhao Qing Song, Shan Zhi Yong. Design and Implementation of Secure Linux Kernel Security Functions. Journal of Computer Research and Development, 2001, Vol.38, No.10, 1255-1261.(in Chinese)
[49] Zhiyong Shan, Tzi-cker Chiueh, Xin Wang. Duplication of Windows Services. CoRR, 2016.
[50] Zhiyong Shan. Suspicious-Taint-Based Access Control for Protecting OS from Network Attacks. Technical Report, 2014.
[51] Zhiyong Shan, Bin Liao. Design and Implementation of A Network Security Management System. Technical Report, 2014.
[52] Zhiyong Shan. A Study on Altering PostgreSQL From Multi-Processes Structure to Multi-Threads Structure. Technical Report, 2014. (in Chinese)
[53] Zhiyong Shan. Implementing RBAC model in An Operating System Kernel. Technical Report, 2015. (in Chinese)
[54] Zhiyong Shan. A Hierarchical and Distributed System for Handling Urgent Security Events. Technical Report, 2014. (in Chinese)
[55] Zhiyong Shan. An Review On Thirty Years of Study On Secure Database and It's Architectures. Technical Report, 2014. (in Chinese)
[56] Zhiyong Shan. An Review on Behavior-Based Malware Detection Technologies on Operating System. Technical Report, 2014. (in Chinese)